\documentclass[preprint,endfloats*]{revtex4}
\nonstopmode

\usepackage{amsmath}
\usepackage{amsbsy}

\newcommand{\ket}[1]{\ensuremath{|#1\rangle}}
 
\newcommand{\dirint}[3]{\ensuremath{\langle #1|#2|#3\rangle}}

\newcommand{\half}{{\textstyle\frac{1}{2}}}

\newcommand{\bs}{\boldsymbol}

\begin{document}

\title{Reply to ``Comment on 'Critique of the foundations of time-dependent density 
functional theory' ''}
\author{J. Schirmer}
\affiliation{Theoretische Chemie, Physikalisch-Chemisches Institut,
Universit\"at Heidelberg, \\
D-69120 Heidelberg, Germany}
\author{A. Dreuw}
\affiliation{Institut f\"ur Physikalische und Theoretische Chemie,
Universit\"at Frankfurt, Germany\\
D-60439 Frankfurt}
\date{\today}

\begin{abstract}
In a recent Comment (arXive:01710.0018), Maitra, Burke, and van Leeuwen (MBL) attempt to refute
our criticism of the foundations of TDDFT (see Phys.~Rev.~A~{\bf 75},~022513~(2007)). 
However, their arguments miss the essence of our position.
This is mainly due to an ambiguity concerning the meaning of the so-called mapping
derivation of time-dependent Kohn-Sham equations. We distinguish two different
conceptions, referred to as potential-functional based fixed-point iteration (PF-FPI)
and direct Kohn-Sham potential (DKSP) scheme, respectively. We argue that 
the DKSP scheme, apparently adopted by MBL, is not a density-functional method at all. 
The PF-FPI concept, on the other hand, while legitimately predicated on the 
Runge-Gross mapping theorem, is invalid because the 
convergence of the fixed-point iteration is not assured.
\end{abstract}
\maketitle
\newpage

Recently we have shown~\cite{sch07:022513}
that time-dependent 
density functional theory (TDDFT) is lacking a valid foundation. In a recent
Comment (arXive:0710.0018), Maitra, Burke, and van Leeuwen (henceforth referred to as MBL) claim to have found
a fault in our argumentation. It seems, however, that  
due to an ambiguity concerning the so-called mapping derivation of the time-dependent 
Kohn-Sham equations
MBL have failed to fully appreciate the essence of our objections to the TDDFT foundations. 
As a matter of fact, their reasoning
can partly be seen as an unintended confirmation of our position rather than a 
refutation, as we will explain in the following.

Before addressing the central issue, let us rectify two minor misunderstandings
brought up in the 2nd paragraph of the Comment, concerning the ``apparent criticisms''
of (i) the Runge-Gross (RG) action-integral functionals, and (ii) the use of non-local
(external) potentials in DFT or TDDFT. The latter point, being the topic of    
Secs.~2 and 3 of our paper, is not at all meant as a critique
of DFT (or TDDFT). On the contrary, this is a highly instructive confirmation of the 
logical consistency of DFT in view of an apparent contradiction.
Second, we fully agree that the failure of the RG action-integral functionals~\cite{run84:997} 
was recognized previously, as has been 
clearly stated in our paper (see Refs. 32 and 36). However, the breakdown of the 
original RG foundation of the time-dependent (TD) Kohn-Sham (KS) equations, 
assuming a stationarity principle for these 
functionals, is not widely known outside the inner TDDFT community. Therefore we have put
particular emphasis on this point in our paper. Accepting this finding breaks the ground 
for the obvious question: How can TD KS equations be established 
without recourse to the RG action integrals?

To a certain extent the misconception 
of our arguments seems to be due to the fact 
that MBL do not always distinguish clearly between
potential-functionals and potentials (as one should not confuse a function with the value of 
that function for a given argument). 
A potential-functional, such as the 
Kohn-Sham potential-functional~(PF),
\begin{equation}
\label{eq:kspf} 
v_{KS}: \rho(\bs{r},t) \rightarrow   v_{KS}[\rho(t)](\bs{r},t)
\end{equation}
(referred to as $w[\rho(t)](\bs{r},t)$ in our paper, see Eq.~75), is a prescription
to construct a (time-dependent) potential $v(\bs{r},t)$ for a given density, $\rho(\bs{r},t)$.
While the KS PF is trivial as a potential-functional, the specific potential
\begin{equation}  
v_{KS}(\bs{r},t) = v_{KS}[\rho_0(t)](\bs{r},t)
\end{equation}
associated with the
exact TD density, $\rho_0(\bs{r},t)$, of the interacting $N$-electron system under consideration
is a highly non-trivial TD potential. Plugging it into the TD KS equation 
for a single orbital (assuming here the radical KS form),
\begin{equation}
\label{eq:kspot} 
i\frac{\partial}{\partial t} \psi(\bs{r},t) = \{-\half \nabla^2 + v_{KS}(\bs{r},t)\} \psi(\bs{r},t)
\end{equation}
would allow one to
determine the time-evolution of the exact one-particle density, $\rho_0(\bs{r},t)$.
However, the question is not if such a potential exists and, when available, could be used to
solve the TD KS equation in the familiar way (that is by time-propagation). 
Indeed, the existence of $v_{KS}(\bs{r},t)$ is almost a triviality, as was demonstrated in our paper
for the rKS case.  
The actual problem is how to determine $v_{KS}(\bs{r},t)$ (and thereby $\rho_0(\bs{r},t)$)
without making use of 
information obtained by somehow solving the full TD $N$-electron Schr\"{o}dinger equation.
Obviously, here the KS PF~(\ref{eq:kspf}) is of no help. We have called it
trivial, because it is completely unspecific (void of physical significance)
and applies to any given TD density.  
The information necessary to generate 
the TD potential associated with a given TD density (trajectory), $\rho(t)$, has to 
be extracted from the density trajectory itself. 
To determine the respective potential at a given time $t$, $\rho(\bs{r},t)$, 
$\dot\rho(\bs{r},t)$, and
$\ddot\rho(\bs{r},t)$ are needed, as we have shown in Sec. 5 of our paper.  

So what is the present foundation of TD KS equations, that is, as 
a density-based method to determine the density $\rho_0(t)$ of the specific
interacting $N$-electron system? The procedure offered here is based 
on the first Runge-Gross (RG1) mapping theorem~\cite{run84:997}, establishing a  
a fixed-point iteration scheme for the desired density  $\rho_0(t)$. 
This derivation is completely  
analogous to the mapping foundation of the KS equations of static DFT, based only 
on the first
Hohenberg-Kohn (HK1) theorem~\cite{hoh64:864}.
The TD KS approach essentially rests on the following three equations
(adopting for simplicity the rKS formulation):
\begin{equation}
\label{eq:ntrivtd}
v_{KS}[\rho(t)](\bs{r},t) \equiv  v_{ext}[\rho(t)](\bs{r},t) + J[\rho(t)](\bs{r},t) 
+ v_{xc}[\rho(t)](\bs{r},t)
\end{equation}  

\begin{equation}
\label{eq:tdKSx}
i\frac{\partial}{\partial t} \psi(\bs{r},t) = 
\{-\half \nabla^2 + u(\bs{r},t) + J[\rho(t)](\bs{r},t) + v_{xc}[\rho(t)](\bs{r},t)\} \psi(\bs{r},t)
\end{equation} 

\begin{equation}
\label{eq:dens}
\rho(\bs{r},t) = N | \psi(\bs{r},t)|^2
\end{equation}

The first equation (Eq.~81 in our paper) serves as a definition of the (non-trivial) 
xc potential-functional. Here
$v_{ext}[\rho(t)](\bs{r},t)$ is the potential-functional established by 
RG1 for $N$ interacting electrons. For a given TD density, $\rho(\bs{r},t)$,
this PF yields the TD potential recovering $\rho(\bs{r},t)$ when 
used in the interacting $N$-electron TD Schr\"{o}dinger equation. In particular,
\begin{equation}
v_{ext}[\rho_0(t)](\bs{r},t) = u(\bs{r},t)
\end{equation} 
where $u(\bs{r},t)$ is the external potential
of the system under consideration.  
In a similar way the (trivial) 
KS potential-functional, $v_{KS}[\rho(t)](\bs{r},t)$, established by the RG1 mapping 
for the non-interacting
KS system (or directly according to Eq.~75 of our paper),  
yields a potential that will recover $\rho(\bs{r},t)$ when used in the TD KS equation.
$J[\rho](\bs{r},t)$ denotes the Hartree potential-functional.

By using the non-trivial xc PF in the KS equation (\ref{eq:tdKSx}) one obtains a
fixed-point iteration scheme with the desired $\rho_0(\bs{r},t)$ as fixed-point. This 
becomes particularly transparent by expressing $v_{xc}[\rho(t)]$ 
according to Eq.~(\ref{eq:ntrivtd}):        
\begin{equation}
\label{eq:tdKSy}
i\frac{\partial}{\partial t} \psi(\bs{r},t) = 
\{-\half \nabla^2 + v_{KS}[\rho(t)](\bs{r},t) +  u(\bs{r},t) - 
v_{ext}[\rho(t)](\bs{r},t) \} \psi(\bs{r},t)
\end{equation} 
If (and only if) $\rho(\bs{r},t) \equiv \rho_0(\bs{r},t)$, the two latter potentials 
on the r.h.s.  
cancel each other, and the KS equation with the remaining potential 
$v_{KS}[\rho_0(t)](\bs{r},t) = v_{KS}(\bs{r},t)$ 
will yield the fixed-point density $\rho_0(\bs{r},t)$.

However, as it stands, this is just an \emph{ad hoc} fixed-point iteration scheme. To qualify as 
a method for determining $\rho_0(\bs{r},t)$ the question of convergence must be settled, 
at least in principle. But so far there is no proof of the possibility of convergence. 
So even if one had the exact xc PF, $v_{xc}[\rho(t)]$, this would be of no avail to 
determine the fixed-point solution, because one cannot expect the fixed-point iteration (referred to 
as trajectory mode solution in our paper) to converge. The situation here
is completely different from the case of static DFT, where the fixed-point iteration is the standard
method to determine the ground-state density $\rho_0(\bs{r})$. In static
DFT, by contrast, the variational principle according to the second Hohenberg-Kohn 
theorem (HK2) assures the
convergence of the fixed-point iteration as a means of finding the (existing) minimum of the
energy functional.

In the TDDFT community, the lack of a proof of convergence for the TD KS 
fixed-point iteration has never been
seen as a problem. To the best of our knowledge, the problem
has not even been mentioned in the TDDFT literature. There was no reason to worry
about the fixed-point iteration at all,  
because there was an apparent silver bullet: direct 
time-propagation of the TD KS equation (\ref{eq:tdKSx}). By starting from the exact static
ground-state density, $\rho_0(\bs{r})$, at the onset of the TD interaction, say at $t=0$, 
it was believed that one could propagate the initital KS orbital through a given time
by supplying the required density argument in $v_{xc}[\rho(t)](\bs{r},t)$ 
(and $J[\rho(t)](\bs{r},t)$) ``on the fly'' via
Eq.~(\ref{eq:dens}).
While this would work for instantaneous potential-functionals 
(as those used in the adiabatic approximation), the potential-functionals in general cannot be
expected to be instantaneous. As we have explicitely shown in our paper, 
the KS potential-functional 
$v_{KS}[\rho(t)](\bs{r},t)$ required in the definition (\ref{eq:ntrivtd}) of 
$v_{xc}[\rho(t)](\bs{r},t)$ depends
on the second time-derivative of the density, which undermines the possibility of propagation.  
The eventuality that the potential-functionals used in the TD KS equations are non-instantaneous
in the described sense has never been seen and addressed before. 
The inescapable conclusion of that finding is that
the TD KS equations as established by the RG1 mapping theorem do not represent
physical equations-of-motion, which could be treated by time-propagation. The silver bullet
is an illusion.

Let us emphasize once again: the dependence on the 
second time-derivative of
the density trajectory (and thus on the infinitesimal ``past'' and ``future'' of the 
density at a given time $t$) pertains to the potential-functionals, $v_{KS}[\rho(t)]$, and
$v_{xc}[\rho(t)]$. We do \emph{not} claim that this  
finding applies to the KS potential, $v_{KS}(\bs{r},t)$, 
itself, as MBL erroneously assume. 
As a matter of fact, $v_{KS}(\bs{r},t)$ can be
determined directly, but only by somehow solving the full TD Schr\"{o}dinger equation (FTDSE) for 
the original $N$-electron system. We do not at all deny that such a direct FTDSE based 
determination 
of $v_{KS}(\bs{r},t)$ can do perfectly well without second time-derivatives of the 
wave function or the density. 
Most of the Comment by MBL is devoted to demonstrate just that.  
 
As was pointed out in our paper, the rKS formulation allows for an obvious direct approach to  
$v_{KS}(\bs{r},t)$ by
solving the FTDSE (e.g. by propagation starting at $t=0$), thus yielding the 
exact TD density, $\rho_0(\bs{r},t)$, which in turn can be inserted as argument
in the explicitely available (trivial) KS potential-functional,
$v_{KS}(\bs{r},t) =  v_{KS}[\rho_0(t)](\bs{r},t)$. 
(This way to determine $v_{KS}(\bs{r},t)$ from  $\rho_0(\bs{r},t)$
is also referred to as the inversion of the TD KS equation.) 
In the final inversion step, being based only on the density information and 
a potential-functional, the 2nd time-derivative of the density is required, which, of course,
causes no problem here, as the density trajectory is available anyway, along with its
first and second time-derivatives. 

At this point MBL think they have refuted us, namely, by demonstrating 
that a direct (FTDSE based) determination of $v_{KS}(\bs{r},t)$ is possible without
recourse to second derivatives, that is, depending only on the past. But this misses our
point completely. The requirements of a direct (FTDSE based) approach to $v_{KS}(\bs{r},t)$          
are completely irrelevant to our criticism of the validity of predictive TD KS equations
according to Eqs.~(\ref{eq:ntrivtd} - \ref{eq:dens}). The real issue  
are the properties of the involved potential-functionals: What is actually needed to generate
a potential at a time $t$ via a potential-functional? As we have shown explicitely: 
the density at $t$, together with its first and second time-derivative at that 
time.

Here we are closing in to the core of the controversy. Apparently,
MBL see the essence of TDDFT in Eq.~(\ref{eq:kspot}) together with the possibility to
determine the KS potential, $v_{KS}(\bs{r},t)$, in a direct way. But it should be clear
that such a direct approach to $v_{KS}(\bs{r},t)$ is, perhaps, 
an interesting detour to determine $\rho_0(\bs{r},t)$  by ultimately solving 
the full TD Schr\"{o}dinger equation, but certainly not a density-functional method.     
Unfortunately, the distinction between the latter direct KS potential scheme and the
genuine density-functional approach (Eqs.~\ref{eq:ntrivtd} - \ref{eq:dens})
is rather blurred than clarified in the TDDFT literature. 
Astonishingly, there is no such thing as a second founding paper 
(of the theorem-proof type), filling the void
after the breakdown of the original RG foundation. Even more disconcerting is
the instance that the PF-FPI procedure, 
while well-known in the inner TDDFT community, 
has not been presented in the literature so far. 
What is offered instead (see, for example, the 
review article by Marquardt and Gross~\cite{mar04:427}) is rather 
misleading. It is insinuated that a non-trivial xc PF can be obtained by just subtracting 
the Hartree PF and the given external potential, $u(\bs{r},t)$, from the trivial KS PF. 
While 
the TD KS equations then assume the familiar, reassuring shape 
of physical 
equations-of-motion, apparently amenable to time-propagation after making a good guess for 
the xc PF, the real issue of establishing a density-based method for td systems 
has got out of sight.

Finally let us take a look at the specific derivations presented in the Comment. Here
MBL demonstrate the possibility to construct the exact xc potential, $v_{HXC}(\bs{r}, t)$
in their notation, directly without recourse to second time-derivatives of the TD density.
But as can readily be seen, this direct approach
is predicated on solving the FTDSE. 
Here the key is to note that 
the Sturm-Liouville type Eq.~C(3) of the Comment,
underlying the construction of $v_{HXC}(\bs{r}, t)$, depends explicitly on
the quantity (Eq.~C2)  
\begin{equation}
\nonumber
q(\bs{r}, t) = \dirint{\Psi(t)}{\hat{q}(\bs{r})}{\Psi(t)}
\end{equation}
that is, an expectation value involving the
the $N$-electron TD wave function $\Psi(t)$.  
This makes apparent that the construction of the xc potential at a 
given time $t$ along Eqs.~C(1) to~C(4)
depends on the TD ($N$-electron) wave function, $\Psi(t)$ at $t$. 

Since according to Eq.~C(3) $v_{HXC}(\bs{r}, t)$ (at a given time $t$) 
is completely determined by $\Psi(t)$ (and by the density $n(t)$, 
being itself determined by $\Psi(t)$), one can readily devise a    
time-propagation scheme for $v_{HXC}(\bs{r}, t)$ based on time-propagation of $\Psi(t)$. 
The time-propagation of $v_{HXC}(\bs{r}, t)$ can, in turn, be used in the propagation  
of TD KS equations. Like the original time-propagation of $\Psi(t)$,
the (simultaneous) time-propagations of $v_{HXC}(\bs{r}, t)$ and the TD KS equations
are viable procedures, depending only on the previous (past) time steps. 
The actual route taken in the demonstration along Eqs.~C(5-11) is slightly different, though.
Rather than using time-propagation for  $\Psi(t)$, the procedure adopted by MBL amounts
to expanding $\Psi(t)$ in terms of a Taylor series at $t=0$.
The first step, spelled out in Eq.~C(8),
establishes $\partial _t \Psi(\Delta t) = -i \hat{H}(0)\Psi (0)$, which readily leads to the
commutator expression on the rhs of Eq.~C(8). 
In the general ($k$th) time step, $q(\bs{r}, t)$ can be written in a form
involving  
multiple commutators of $\hat{q}(\bs{r})$ and $\hat{H}(0)$ sandwiched
by $\ket{\Psi_0} = \ket{\Psi(0)}$ (see text after Eq.~C11). Obviously,
the multiple commutators arise from terms of the kind $\hat{H}(0)^n \ket{\Psi_0}$
with increasingly higher powers of $\hat{H}(0)$. 

Let us note that the Taylor-type expansion of $\Psi(t)$
means a severe restriction with respect to genuine time-propagation. 
The Taylor representation (at $t=0$) is  valid only for a
time-independent Hamiltonian and, moreover, presupposes that the wave function
is analytic in time at $t=0$. (Already the 2nd time derivative of the wave function
is discontinuous at $t=0$ for a simple TD potential of the form $ \sim \Theta(t)\, t$.)

In any case, what MBL have shown here is that the exact xc potential can be constructed
directly and used in a viable time-propagation scheme of the TD KS equations - if
the $N$-electron wave-function $\Psi(t)$ is at the disposal. 
As explained above, 
we have no objections against this demonstration, it just misses the point.

As our Reply has grown somewhat long, it may be helpful to summarize the essentials.
1)~After the breakdown of the original foundation of TDDFT based on the RG
action-integral functionals and their stationarity principles, there is a confusion
about the essence of the alternative mapping derivation of TD KS equations, resting on
the RG1 mapping theorem. The authors of the Comment seem to adhere to an idea denoted
as direct KS potential (DKSP) scheme here. By contrast, there is a more common view 
that the TD KS equations have to be established as a fixed-point iteration scheme for 
potential-functionals (PF-FPI), following here an analogous approach in static DFT.
2)~In our paper we have shown that the PF-FPI concept is not valid because, by contrast to 
static DFT, the convergence of the fixed-point iteration is not assured. 
The apparent way out, that is,
time-propagation of the TD KS equations (in the PF-FPI notion) 
fails because the potential-functionals require the second time-derivative
of their density arguments. Our criticism of the PF-FPI concept has not been addressed, let alone
refuted by MBL, focussing rather on the DKSP scheme. 
3)~We argue that the DKSP scheme is not a density-functional method at all (i.e. based entirely on 
densities and density potential-functionals), because there is no way to determine 
the required KS potential, $v_{KS}(\bs{r},t)$, other than by somehow solving (or approximating)
the full (interacting $N$-electron) TD Schr\"{o}dinger equation. 
As an unintended confirmation of this point one may see the large passage in the Comment
where MBL show that the DKSP equation can be propagated without the need for 
second-time derivatives: unwittingly
they here make use of the time-propagation of the full $N$-electron wave function.

In conclusion, the status of TDDFT as an in principle rigorous approach to 
the time-evolution
of an interacting many-electron system, must still be considered as completely unfounded.
Instead of advocating ``many more exciting real-life applications'' of TDDFT, 
a major effort should be made
to ultimately clarify the question whether one can derive predictive equations-of-motion
formulated entirely in terms of densities, that is, without recourse to the wealth 
of phase information 
guiding the time-evolution of the many-electron wave functions. 
It is highly questionable if one will be able
to derive density-based physical equations-of-motion at all, but it will be next to impossible
to do so without a valid variational (or stationary) principle.


\newpage
\begin{thebibliography}{4}
\expandafter\ifx\csname natexlab\endcsname\relax\def\natexlab#1{#1}\fi
\expandafter\ifx\csname bibnamefont\endcsname\relax
  \def\bibnamefont#1{#1}\fi
\expandafter\ifx\csname bibfnamefont\endcsname\relax
  \def\bibfnamefont#1{#1}\fi
\expandafter\ifx\csname citenamefont\endcsname\relax
  \def\citenamefont#1{#1}\fi
\expandafter\ifx\csname url\endcsname\relax
  \def\url#1{\texttt{#1}}\fi
\expandafter\ifx\csname urlprefix\endcsname\relax\def\urlprefix{URL }\fi
\providecommand{\bibinfo}[2]{#2}
\providecommand{\eprint}[2][]{\url{#2}}

\bibitem[{\citenamefont{Schirmer and Dreuw}(2007)}]{sch07:022513}
\bibinfo{author}{\bibfnamefont{J.}~\bibnamefont{Schirmer}} \bibnamefont{and}
  \bibinfo{author}{\bibfnamefont{A.}~\bibnamefont{Dreuw}},
  \bibinfo{journal}{Phys. Rev. A} \textbf{\bibinfo{volume}{75}},
  \bibinfo{pages}{022513} (\bibinfo{year}{2007}).

\bibitem[{\citenamefont{Runge and Gross}(1984)}]{run84:997}
\bibinfo{author}{\bibfnamefont{E.}~\bibnamefont{Runge}} \bibnamefont{and}
  \bibinfo{author}{\bibfnamefont{E.~K.~U.} \bibnamefont{Gross}},
  \bibinfo{journal}{Phys. Rev. Lett.} \textbf{\bibinfo{volume}{52}},
  \bibinfo{pages}{997} (\bibinfo{year}{1984}).

\bibitem[{\citenamefont{Hohenberg and Kohn}(1964)}]{hoh64:864}
\bibinfo{author}{\bibfnamefont{P.}~\bibnamefont{Hohenberg}} \bibnamefont{and}
  \bibinfo{author}{\bibfnamefont{W.}~\bibnamefont{Kohn}},
  \bibinfo{journal}{Phys. Rev.} \textbf{\bibinfo{volume}{136}},
  \bibinfo{pages}{B 864} (\bibinfo{year}{1964}).

\bibitem[{\citenamefont{Marques and Gross}(2004)}]{mar04:427}
\bibinfo{author}{\bibfnamefont{M.~A.~L.} \bibnamefont{Marques}}
  \bibnamefont{and} \bibinfo{author}{\bibfnamefont{E.~K.~U.}
  \bibnamefont{Gross}}, \bibinfo{journal}{Annu. Rev. Phys. Chem.}
  \textbf{\bibinfo{volume}{55}}, \bibinfo{pages}{427} (\bibinfo{year}{2004}).

\end{thebibliography}
\end{document}